\documentclass[twocolumn,superscriptaddress,pra,a4paper]{revtex4-1}
\usepackage{graphicx,float}

\usepackage[utf8]{inputenc} 

\begin{document}

\title[Construction biases in biological databases]{Detection of construction biases in biological databases: the case of miRBase}
 
\author{Guilherme Bicalho Saturnino}\email{guilherme.bicalho2@gmail.com}
\affiliation{Departamento de Física, Universidade Federal de Minas Gerais, Belo Horizonte, MG, Brazil}
\author{Caio Padoan de Sá Godinho}\email{godinhocps@gmail.com}
\affiliation{Departamento de Ciências Biológicas, Universidade Federal de Ouro Preto, Ouro Preto, MG, Brazil}
\author{Denise Fagundes-Lima}\email{defalima@gmail.com}
\affiliation{Departamento de Física, Universidade Federal de Minas Gerais, Belo Horizonte, MG, Brazil}
\author{Alcides Castro e Silva}\email{alcidescs@gmail.com}
\affiliation{Departamento de Física, Universidade Federal de  de Ouro Preto, Ouro Preto, MG, Brazil}
\author{Gerald Weber}\email{gweberbh@gmail.com}
\affiliation{Departamento de Física, Universidade Federal de Minas Gerais, Belo Horizonte, MG, Brazil}

\begin{abstract}
Biological databases can be analysed as a complex network which may reveal some its underlying biological mechanisms.
Frequently, such databases are identified as scale-free networks or as hierarchical networks depending on 
connectivity distributions or clustering coefficients. 
Since these databases do grow over time, one would expect that their network topology may undergo some changes. 
Here, we analysed the historical versions of miRBase, a database of microRNAs
where we performed an alignment of all mature and precursor miRNAs and calculated a pairwise similarity index.
We found that the clustering coefficient shows important changes during the growth of this database.
For two consecutive versions of the year 2009 we found a strong modification of the network topology which
we were able to associate to a technological change in miRNA discovery.
To evaluate if these changes could have happened by chance, 
we performed a set of simulations of the database growth by sampling the final version of miRBase
and creating several alternative histories of miRBase.
None of the simulations were close to the actual historical evolution of this database, which we
understand as a clear indication of a very strong construction bias.
\end{abstract}
\maketitle

Since Barabasi's seminal work~\cite{barabasi99} the concept of complex networks has spread to 
nearly all fields of research. 
Countless systems can be modelled as complex networks, for instance in social interactions 
(human relationship, inter-city and intra-city movements~\cite{apicella2012social}), 
ecology (predation, mutualism~\cite{montoya2006ecological,bascompte09}), transport (traffic flow, airports~\cite{barrat2004architecture}).
Even food recipes can be analysed in this way~\cite{ahn2011flavor}.
Unsurprisingly, networks built from interactions in molecular biology were among the first
applications of complex network analysis and have since become a standard part of bioinformatics textbooks, 
see for example
Ref.~\onlinecite{lesk08}.
One of the best known examples of networks in molecular biology are protein interaction networks
where the network is formed with proteins that bind to each other in 
order to carry out some biological function~\cite{blow09,lees11}.
Also, gene regulatory networks where genes interact via transcription factors and the metabolic network~\cite{pan07}
are examples of network topology analysis. 
The usage of biological networks has become so common that even a word was invented to describe more 
synthetically the whole set of interactions in a cell: the interactome.  

The growing popularity of network topology analysis in molecular biology is easily explained in terms of the promise
to uncover fundamental new biological insights while analysing massive amounts of data.
Yet one key question appears to have been systematically overlooked: what if the network topology is not 
of biological origin but the consequence of particular technological constraints by which the database 
was populated with biological data?
To stay within the example of well known protein interaction networks, suppose that some network analysis shows that 
a given protein is found to form a hub of highly connected interactions. 
One possibility would be that this particular protein simply attracted more research activity resulting in more entries in the 
database while other proteins received less attention.
Clearly, network topology analysis could be entirely skewed by biases in the construction of the database.
To our knowledge, this problem has not seen attention or being acknowledged.
Note that we are not discussing in this work the specific issue of scale-free topologies of biological networks
which were found to be problematic~\cite{tanaka05b,prvzulj04,khanin06}.

The main problem here is that biological databases are inherently, but not uniformly, incomplete.
The present day database is essentially a subset of what biology has in store.
Therefore database construction biases can be regarded as a problem of subnet sampling.

Sampling a part of a large network is a common and mostly unavoidable procedure especially
for very large networks.
For example it is impossible to analyse the network topology of the entire internet since many 
network quantities scale to the square or even to the cube of the number of nodes.
This problem results in the following classical question: does the subnet share the same topological
properties of the whole net?
Evidently, this is a difficult question which attracted some attention,
sometimes with seemingly contradicting findings~\cite{stumpf05,han05}. 
One definite conclusion though is that the resulting topology is crucially dependent on the sampling
strategy and that the property of clustering appears to be the most sensitive to sampling~\cite{lee06c}.

For the question we wish to analyse, if there is a database construction bias, there is no choice of sampling 
strategy available.
The subnet sampling is entirely determined by whatever data was obtainable by current technical methods at a certain period
of time.
Consider for example, a network which grows over time.
Any previous version of this network is a subnet of the newer version.
Clearly, in this case the network topology could change over time.
For instance, a network could start out as random in its early days and gradually mature into 
a scale free network if some preferential attachment is involved.

The question therefore is not only if there is a construction bias at all, but how much data is necessary for this bias
to become negligible.
One aspect of database formation makes this question hard to answer: there are many levels of information in a database and 
all of them depend to some extent on annotation.
For example, a database may contain a specific genomic sequence and also if this sequence is related in some way
to other sequences.
To form a network one usually combines multiple informations to connect different nodes which means that 
the network topology results from an entangled mix of information.

Here we will attempt to address the detection of database construction biases by using the smallest amount of information
possible.
To achieve this we selected one relatively small yet important database: the database of microRNA (miRNA)
known as miRBase~\cite{griffiths-jones06}.
The discovery of miRNAs is fairly recent and the database has seen a peak of activities over the 
last years~\cite{kozomara11}.
Also, the discovery of new miRNAs has improved considerably since the early days of this database.
We can now ask our question again for this database: is the network topology of this database biased
by the way new miRNAs are discovered? 
If yes, which topology parameters would be affected and how strongly?

To form a network of miRNAs we need to chose how to link them. 
There are many possibilities, for example we could link them by their silencing targets, by the genes which express 
them or a mixture of different criteria.
The problem with this is approach is that the link would depend on multiple layers of database annotation and would be a source
of complication for our analysis.
Instead, we selected to establish this relation with sequence similarities which
has the advantage of being a totally deterministic way of establishing a relation between two miRNAs.
Therefore, our network depends solely on one information contained in the database which is the miRNA sequence and 
nothing else.
Another motivation is that some of the methods for discovering new miRNAs rely on sequence similarities and therefore
this may become apparent in the network topology. 

A further reason which makes miRNAs attractive for similarity network analysis is their short length, around 20~nt 
for mature miRNAs and of the order of 100~nt for precursor sequences. 
It is their short lengths which allowed us to calculate the pairwise sequences similarities 
for the complete miRBase.
We calculated the sequence similarities of all miRNA sequences compared to each other which provides us with
a similarity index. 
Then we calculated the network topology parameters such as average weights
and clustering coefficients for each
of the historical versions of miRBase.
We performed a series of simulations which basically ask the following question: what if history had
happened differently? What if the miRNAs were discovered in a different order? Can we actually distinguish
the real database from the simulated one?
By analysing a subset of the final database we are essentially borrowing robust statistics or bootstrapping methods. 
It is
common to chose randomly a subset of a larger set of data and then estimate how much a given parameter
deviates when compared to the original set~\cite{press88}.
For instance, in a recent work~\cite{weber09b} we were able to estimate error bars for DNA flexibility using
this type of method.

The main rationale behind our approach is that if sequences were deposited in a database in a certain biased order
any resulting network topology, no matter which linking criteria was used to form the network, will be biased as well.
Note that an unbiased database should result in similar network topology parameters over all 
database versions. All topology parameters should confirm this.
If just one parameter presents a bias this establishes unambiguously that the dataset history is biased
and no further tests are necessary.
Considering that similarity networks also have found their uses for sequence clustering~\cite{miele11}, for multiple alignment of
protein sequences~\cite{miele12}, phylogenetic analysis~\cite{andrade11} and visualization of protein
superfamilies~\cite{atkinson09}, the conclusions presented here may be relevant for these applications.

\section*{Methods}

\subsection*{Similarity index}
We used a standard Needleman-Wunsch (NW)~\cite{needleman70} alignment algorithm to calculate the sequence similarity
of all miRNAs against each other. 
The alignment matrix~$P$ is filled according to the following rule
\begin{equation}
P(i,j)=\max\left\{
\begin{array}{l}
P(i-1,j-1)+R(i,j)\\
P(i-1,j)+g\\
P(i,j-1)+g
\end{array}
\right.
\label{eq-P}
\end{equation}
where $i$ and $j$ are the nucleotide positions of the two sequences which are being compared, and
\begin{equation}
R(i,j)=\left\{
\begin{array}{ll}
m & \hbox{if $i$ and $j$ are the same nucleotide} \\
d & \hbox{otherwise}
\end{array}
\right.
\end{equation}
The alignment matrix~$P$ expresses the degree of similarity between to sequences.
Each cell of the matrix is evaluated sequentially using the rule $R$ and compared against the previous cells.
If the two sequences are very similar the score $m$ is used more frequently producing higher values
for these cells.
In this work we used the score $m=2$ which is used when two nucleotides are identical.
The score $d=-1$ is a penalty when two nucleotides are different, and gaps in the sequences are
penalized even more by $g=-2$.
For sequences of length~$l$ the highest score $S$ that can be achieved is $m^l$, that is, 
the score $m$ ($l$ nucleotides are identical) is summed over $l$ times when evaluating Eq.~(\ref{eq-P}).
We then normalize the NW score to one, which we call the similarity index~$s=S/m^l$.
Therefore, two identical sequences will score $s=1$.

These similarity indices are used to build a weighed network where $w_{ij}$ is the similarity $s$
between miRNAs $i$ and $j$.
One advantage of using the NW algorithm is that it results in symmetrical
weights, that is $w_{ij}=w_{ji}$, therefore avoiding problems with unsymmetrical similarities~\cite{andrade11}
and other errors associated to Blast searches~\cite{gonzalez10}.
For $N$ microRNAs this results into $N(N-1)/2$ similarity relations and to reduce the size of
the networks we remove all links below a minimal score~$w_{ij}< s_m$~\cite{castroesilva08}.
We have chosen a minimal score $s_m=0.4$ that was found to retain all nodes with at least
one connection.
This removal is also necessary since similarity indices below $s\approx0.25$ may occur by chance and
therefore have no real similarity meaning (see also Fig.~1 of Ref.~\onlinecite{castroesilva08}).

\subsection*{Clustering coefficient}
The discrete clustering coefficient is commonly defined as
\begin{equation}
c_i=\frac{1}{k_i(k_i-1)}\sum_{j,h\neq i}a_{ij}a_{ih}a_{jh}
\label{eq-c}
\end{equation}
where the coefficients $a_{ij}$ will be equal to one if there is a 
connection between the nodes $i$ and $j$ and zero otherwise.
The connectivty, that is, the total number of connections leading to node $i$ is $k_i$.
Note that in the similarity network all nodes are connected by a weight $w_{ij}$, therefore the discrete
clustering coefficient only differs from unity if we apply a minimal score as described in the previous section.

The weighted clustering coefficient which is formally closest to the discrete coefficient of Eq.~\ref{eq-c}
is based on the geometric
average proposed by Onnela et al.~\cite{onnela05}
\begin{equation}
c^w_i=\frac{1}{k_i(k_{i}-1)}\sum_{j,h=1}^N(w_{ij}w_{ih}w_{jh})^{1/3}
\end{equation}
where $w_{ij}$ is the normalized weight of the connection between the nodes~$i$ and~$j$,
and $k_i$ is the connectivity of node~$i$. 
Basically this coefficient counts how many triplets a node and its 
neighbours can form.
The coefficient is normalized such that it will be $0$ if there are no triplets
and $1$ if all triplets are maximum. 
The
clustering coefficient is a node-related property and it is common to define a global discrete and weighted clustering coefficient, 
averaging over all $c_i^w$, namely:
\begin{equation}
C=\frac{1}{N}\sum_{i=1}^Nc_i
\end{equation}

\begin{equation}
C^w=\frac{1}{N}\sum_{i=1}^Nc^w_i
\end{equation}

\subsection*{Simulation}
We performed several simulations to determine the database topology if the miRNAs deposited in miRBase were discovered
in a different chronological order.
Two different types of alternative first versions were attempted, either
we sampled randomly from the last version or we started with the same version as the initial database 1.2.
The simulations described in this section were performed for both mature and precursor miRNAs.

\paragraph{Simulation with fixed initial version (FIV)}
We selected all sequences in the initial version 1.2 that do exist up to version (17.0), that is,
the original version 1.2 minus the sequences which were deleted up to the last version. 
In this case, every time we simulate new higher versions we start over with exactly the same initial version.

\paragraph{Simulation with random initial version (RIV)}
Here, the new initial version corresponding to database 1.2 is
randomly populated with sequences from version 17 until we reach the same number of elements as the original 
version 1.2.

\paragraph{Populating the remaining database versions}
The versions after 1.2 of the database are populated randomly and without repetition,
sampling from version 17.0 with the following criteria
\begin{enumerate}
\item first we make a set of sequences in 17.0 which have a similarity index~$s>0.95$ with any 
sequence already present in the database. We then sample uniformly from this set.
\item should we run out of sequences with $s>0.95$ before completing the new version, 
then we sample randomly the remaining sequences from version 17.0.
\end{enumerate}
Once a given version is completely populated and contains the same amount of sequences as the original 
version we perform all network topology calculations. 
For each database version this simulation is carried out 20 times.

This strategy intents to simulate a process by which 
the next version of the database is populated by sequences which are closely related to sequences which already 
exist in previous versions.
In this way we try to mimic the effect that new sequences may be found by similarity to known ones,
that is we try to introduce some level of bias.

\paragraph{Simulation for mature \textit{H. sapiens} miRNAs (HS)}
Here we select only the mature miRNAs for \textit{H. sapiens} which is the largest subset of miRBase and 
rebuild the versions containing the same of \textit{H. sapiens} in each version.

\subsection*{Sequences used}

We used all mature and precursor sequences and versions of miRBase up to version 17.0 corresponding to the
version when this work was started~\cite{griffiths-jones06,kozomara11}.
Table~\ref{tab-mirbase} summarizes all sequences used in this work.

\begin{table}[tbp]
\caption{Summary of the miRBase database versions. Shown are the sequential index, the official
version designation, the date on which this version was released, 
the number of precursor miRNAs $N_{\rm pre}$,
the number of mature miRNAs $N_{\rm mat}$, the number of mature miRNAs which are not
present in the last version $R_{\rm mat}^{17}$,
and the number of mature miRNAs of \textit{H. sapiens} $N_{\rm HS}$.
\label{tab-mirbase}}
\begin{center}
\begin{tabular}{rrcrrrrrr}\hline\hline
index & version & date & species & $N_{\rm pre}$ & $N_{\rm mat}$ & $R_{\rm mat}^{17}$ &$N_{\rm HS}$ \\ \hline 
1 &  1.1 &  01/2003 & 5 & 265 & 265 & 112 & 87 \\
2 &  1.2 &  04/2003 & 5 & 295 & 256 & 103 & 87 \\
3 &  1.3 &  05/2003 & 5 & 332 & 295 & 103 & 87 \\
4 &  1.5 &  07/2003 & 5 & 400 & 370 & 160 & 88 \\
5 &  2.0 &  07/2003 & 6 & 506 & 464 & 171 & 135 \\
6 &  2.2 &  11/2003 & 7 & 593 & 528 & 188 & 136 \\
7 &  3.0 &  01/2004 & 8 & 719 & 644 & 212 & 152 \\
8 &  3.1 &  04/0204 & 8 & 889 & 807 & 242 & 169 \\
9 &  4.0 &  07/2004 & 10 & 1185 & 1143 & 279 & 188 \\
10 & 5.0 &  09/2004 & 12 & 1345 & 1298 & 288 & 189 \\
11 & 5.1 &  12/2004 & 13 & 1420 & 1359 & 257 & 207 \\
12 & 6.0 &  04/2005 & 21 & 1650 & 1591 & 378 & 211 \\
13 & 7.0 &  06/2005 & 33 & 2909 & 2634 & 348 & 313 \\
14 & 7.1 &  10/2005 & 37 & 3424 & 3102 & 366 & 319 \\
15 & 8.0 &  02/2006 & 37 & 3518 & 3229 & 368 & 328 \\
16 & 8.1 &  05/2006 & 39 & 3963 & 3685 & 408 & 455 \\
17 & 8.2 &  07/2006 & 39 & 4039 & 3834 & 418 & 454 \\
18 & 9.0 &  10/2006 & 43 & 4361 & 4167 & 478 & 470 \\
19 & 9.1 &  02/2007 & 43 & 4449 & 4274 & 476 & 470 \\
20 & 9.2 &  05/2007 & 46 & 4584 & 4430 & 494 & 471 \\
21 & 10.0 &  08/2007& 49 & 5071 & 4922 & 284 & 555 \\
22 & 10.1 & 12/2007 & 56 & 5395 & 5234 & 326 & 564 \\
23 & 11.0 & 01/2008 & 62 & 6396 & 6211 & 356 & 677 \\
24 & 12.0 & 09/2008 & 77 & 8619 & 8273 & 408 & 697 \\
25 & 13.0 & 03/2009 & 94 & 9539 & 9169 & 504 & 703 \\
26 & 14.0 & 09/2009 & 103 & 10867 & 10566 & 508 & 718 \\
27 & 15.0 & 04/2010 & 120 & 14197 & 15632 & 641 & 1100 \\
28 & 16.0 & 08/2010 & 129 & 15172 & 17341 & 592 & 1223 \\
29 & 17.0 & 04/2011 & 138 & 16772 & 19724 &     & 1733 \\\hline 
\end{tabular}
\end{center}
\end{table}

\section*{Results and Discussion}

\begin{figure}[tbp]
\begin{center}
\includegraphics[width=6cm]{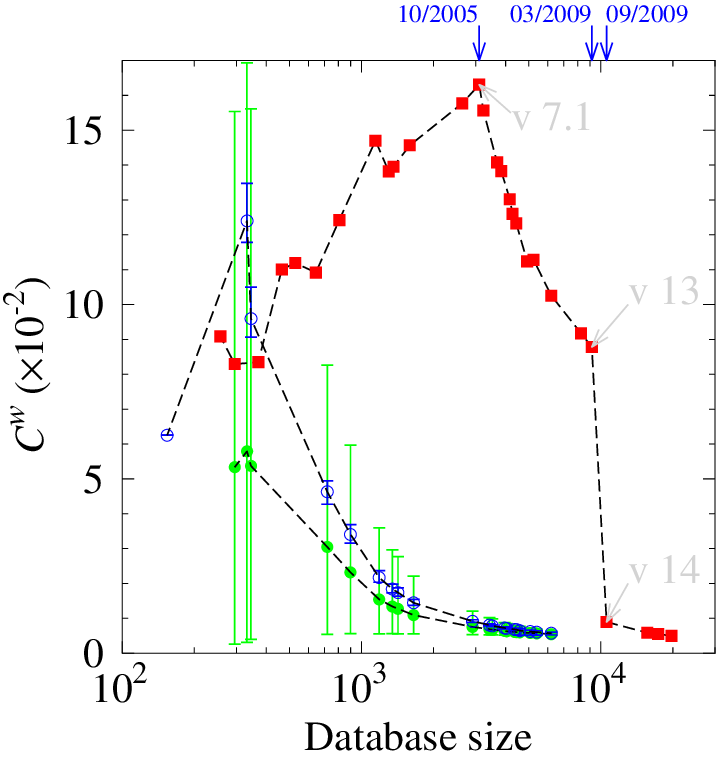}
\end{center}
\caption{
{\bf Average weighted clustering coefficients $C^w$ of mature miRNAs as a function of database size.} 
Red boxes are for the real miRBase versions and blue/green circles are for the simulated database versions.
The simulated database versions are represented by blue circles for a random initial database, and
green boxes if the they start from version 1.2.
}
\label{fig-mature-12}
\end{figure}

In Fig.~\ref{fig-mature-12} we show the average clustering weighted coefficient $C^w$
of mature miRNAs for each version of miRBase. During the initial years the clustering coefficients 
(red boxes)
grow steadily reaching a peak in 2005 from which on they start decreasing again.
In 2009 we observe a dramatic reduction in clustering coefficients which occurs between versions~13 and 14.
We simulated the alternative history of the database with two different starting versions.
In one 
(blue circles)
we started with an entirely random initial version (RIV) which is sampled from the last version~17.
The other one 
(green boxes)
starts with the real version 1.2 (FIV), but then grows randomly from one version to the next.
Neither simulation procedure comes close to reproducing the clustering evolution of the real
database, despite the strong bias introduced in the sampling procedure (see Methods).
In both cases, the clustering coefficients are rapidly dominated by the
low connectivity of the miRNAs which appear since version 14.
From version 13 to the last 17, the database nearly doubles in size which means that at least half of
the miRNAs are sequences with very low clustering coefficient.
This would explain the difficulty in obtaining higher clustering coefficients in any database versions.

\begin{figure*}[bt]
\begin{center}
\includegraphics[width=10cm]{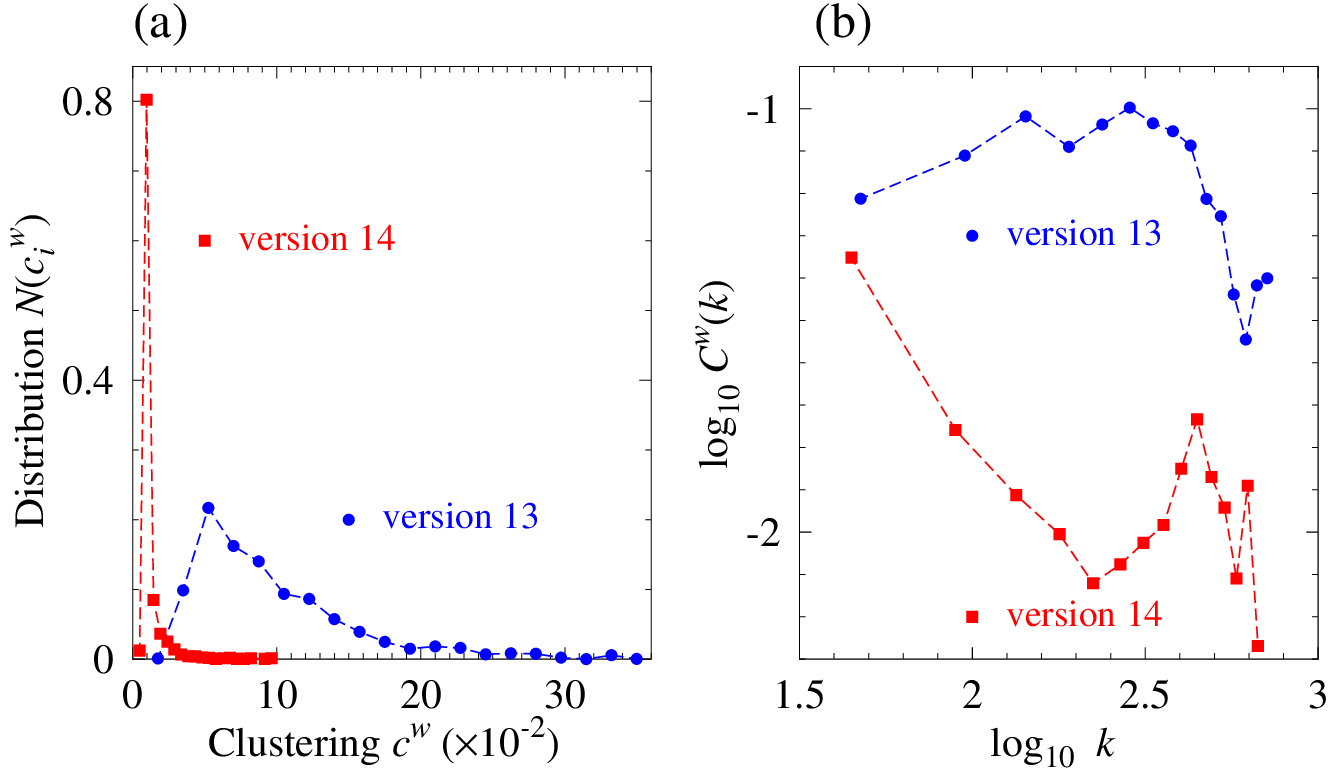}
\end{center}
\caption{
{\bf Clustering coefficient distribution of  mature miRNAs.} 
Shown are (a)~the histograms $N(c_i^w)$ and 
(b)~the average clustering coefficients as function of connectivity $C^w(k)$.
Blue circles are for version~13 and red solid boxes for version 14.
}
\label{fig-coef-dist}
\label{fig-c-k}
\end{figure*}

\begin{figure*}[tbp]
\begin{center}
\includegraphics[width=12cm]{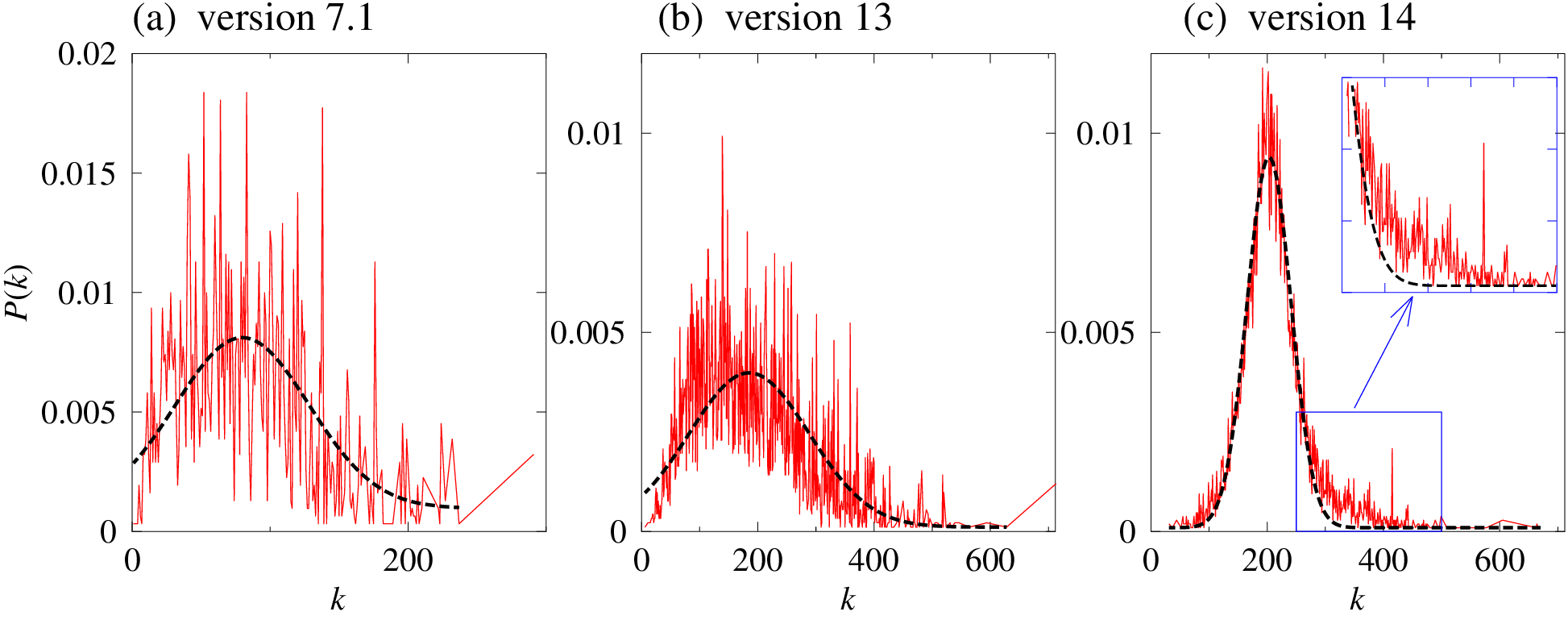}
\end{center}
\caption{
{\bf Connectivity distribution $P(k)$ as a function of connections~$k$ of mature miRNAs.} 
Shown are versions (a)~7.1, (b)~13 and (c)~14.
Solid red curves are the $P(k)$ distributions and dashed black curves are the calculated
symmetrical Gaussian regression of $P(k)$.
}
\label{fig-pk}
\end{figure*}

Fig.~\ref{fig-coef-dist}a shows the difference in clustering distribution between versions~13 and 14.
While the clustering coefficients are distributed over an extended range for version 13, for the next
version they form a delta-like distribution at a very low value.
The distribution of weighted clustering coefficients $C^w$ as function of connectivity~$k$, or $C^w(k)$,  shown in
Fig.~\ref{fig-c-k}b confirms this sudden and dramatic change in network topology.
For version~13, the average clustering coefficients are roughly constants in the $\log-\log$ plot until
$k=400$ ($\log k=2.6$).
In contrast, for version~14, the clustering coefficients shown in Fig.~\ref{fig-c-k}b follow
a power law which resembles that of a hierarchical network until $k=200$ ($\log k=2.3$).
The connectivity distribution~$P(k)$ however, is closer to a random network, as shown in Fig.~\ref{fig-pk}.
The inset of Fig.~\ref{fig-pk}, suggest a slight deviation from a purely random network.


\begin{figure}[tbp]
\begin{center}
\includegraphics[width=7cm]{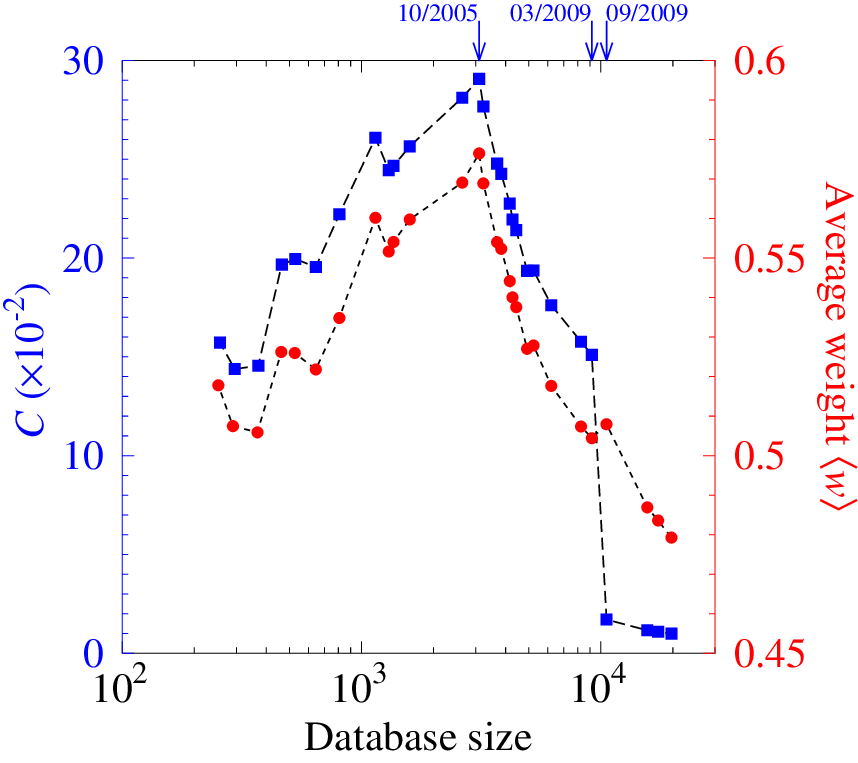}
\end{center}
\caption{
{\bf Average weights $\langle w\rangle$ and discrete global clustering coefficient $C$ of mature miRNAs as a 
function of database size.}
Blue boxes are the average discrete clustering coefficient (left scale) and red bullets 
are the average weights (right scale).
}
\label{avw-mature}
\end{figure}

One question which arises is if this change between versions~13 and~14 is driven by variations in similarity indexes which represent
the weight in our network.
In Fig.~\ref{avw-mature} we show the change of average weight $\langle w\rangle$ with database size (right scale).
Overall we see a very similar behaviour as for the global clustering coefficients $C^w$ with
the notable exception of the sudden drop between versions~13 and~14.
Fig.~\ref{avw-mature} also shows the global discrete clustering coefficients $C^k$ (left scale), that is, 
where the weights were replaced by 1 or 0 depending on the miRNAs being connected or not with a score larger than 0.4.
In this case we see the sudden decrease, which leaves us with the conclusion that it is entirely due to a 
complete change in network connectivity. 
In other words, the new miRNAs deposited in version~14 of miRBase compare to each other with basically the same similarity
index as in version~13, but they do so with much fewer other miRNAs.

\begin{figure}[tbp]
\begin{center}
\includegraphics[width=6cm]{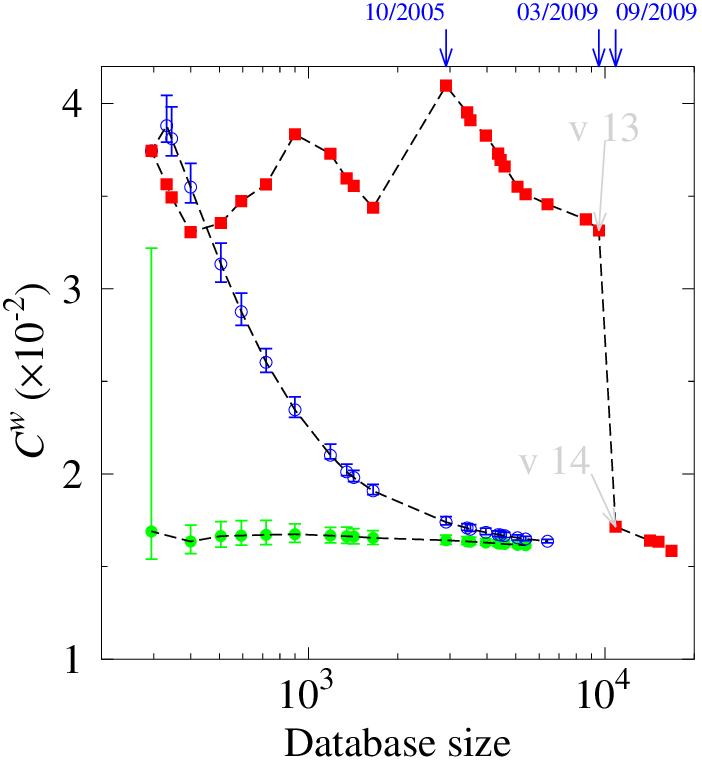}
\end{center}
\caption{
{\bf Global clustering coefficients $C^w$ of pre-miRNAs as a function of database size.} 
Red solid boxes are for the real miRBase versions.
The simulated database versions are represented by blue circles for a random initial database, and
green boxes if the they start from version 1.2.
}
\label{fig-pre-12}
\end{figure}

Precursor miRNAs, that is the longer and unprocessed sequences which results in mature miRNAs, display
a very similar evolution of clustering coefficients shown in Fig.~\ref{fig-pre-12}.
We observe the same peak for version 7.2 as for mature miRNAs, and also the sudden drop between versions 13 and 14.
Overall, the clustering coefficients are lower for the pre-miRNAs than for the mature miRNAs,
which is mainly due to the fact that similarity indexes tend to be much smaller between miRNAs.
Smaller indexes are expected for longer sequences where it becomes less probable that two sequences are similar by
chance.
Nevertheless, the sudden reduction in Fig.~\ref{fig-pre-12} is unmistakable.
As for mature miRNA, the simulations do not reproduce larger clustering coefficients and are rapidly
dominated by the less connected pre-miRNA of later versions of miRBase.
For the simulation which starts with the real initial version 1.2 (FIV), the subsequent versions display some
large clustering coefficients which do not decrease so rapidly.
The simulation with initial random database version (RIV) basically stays constant at the same average clustering
coefficient for all versions.
Only the initial database version shows a very large error bar with nearly disappears for the next versions.

\begin{figure}[tbp]
\begin{center}
\includegraphics[width=6cm]{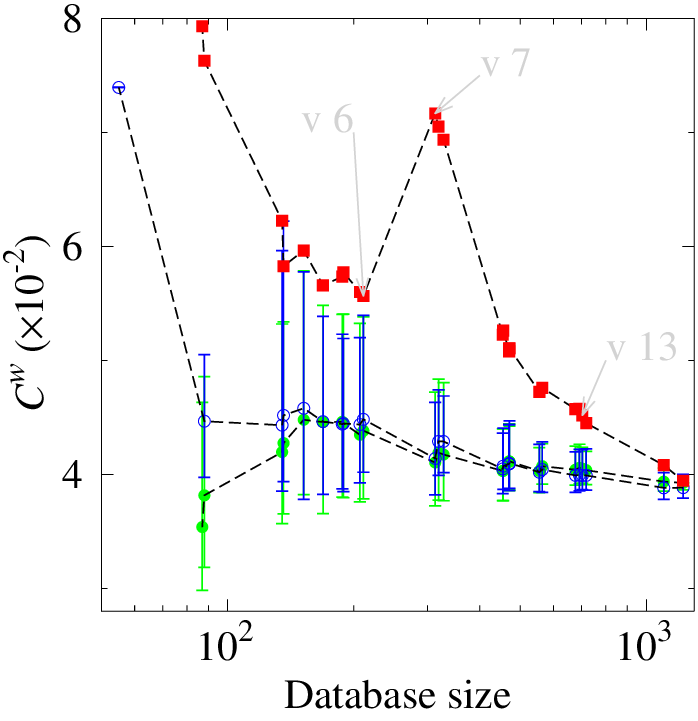}
\end{center}
\caption{
{\bf Average clustering coefficients $C$ of human mature miRNAs as a function of database size.} 
Red solid boxes are for the real miRBase versions.
The simulated database versions are represented by blue circles for a random initial database, and
green boxes if the they start from version 1.2.
}
\label{fig-mature-hs-12}
\end{figure}

Human miRNAs is the most numerous species group present in miRBase, making up about 10\% of the final
version of the database.
We isolated all human mature miRNAs in the database and treated them independently from their remaining  
species groups. 
Their average weighted clustering coefficients are shown in Fig.~\ref{fig-mature-hs-12}.
Unlike the complete mature miRNA network, the clustering coefficient decreases from its initial
version until version~6.
From version~6 to~7 there is a steep increase and a peak, followed by a smooth decrease until the last version of 
miRBase under consideration.
In this case we do not observe the sudden reduction between versions~13 and~14.
While one may argue that there are only 15 new human miRNAs in version~14, no sudden drop is observed for any
later versions even though there is an increase of nearly 400 new sequences between versions~14 and 15.
It is interesting to note that the number of human miRNAs present in miRBase is already larger than some
predicted upper estimate which were made recently~\cite{pritchard12}.
Nevertheless, this means that the set of human miRNAs may be already quite complete.
Yet even in this case, no simulation 
did reproduce the actual course of miRNA discovery.
We have not attempted to calculate the network topology for other species groups in separate, as these
groups are still too small to provide meaningful network topology parameters.

\begin{figure*}[tbp]
\begin{center}
\includegraphics[width=12cm]{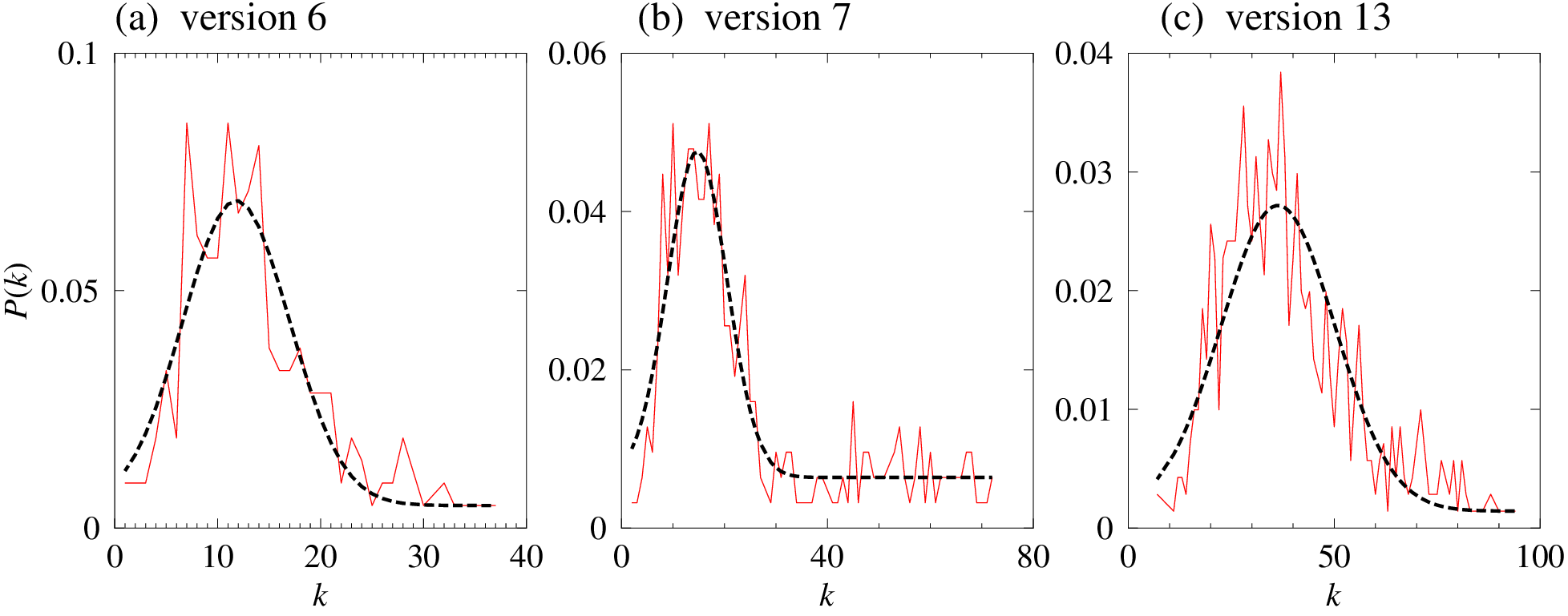}
\end{center}
\caption{
{\bf Connectivity distribution $P(k)$ as a function of connections~$k$ of human mature miRNAs.} 
Shown are versions (a)~6, (b)~11 and (c)~16.
Solid red curves are the $P(k)$ distributions and dashed black curves are the calculated
symmetrical Gaussian regression of $P(k)$.
}
\label{fig-pk-hs}
\end{figure*}

\begin{figure*}[tbp]
\begin{center}
\includegraphics[width=12cm]{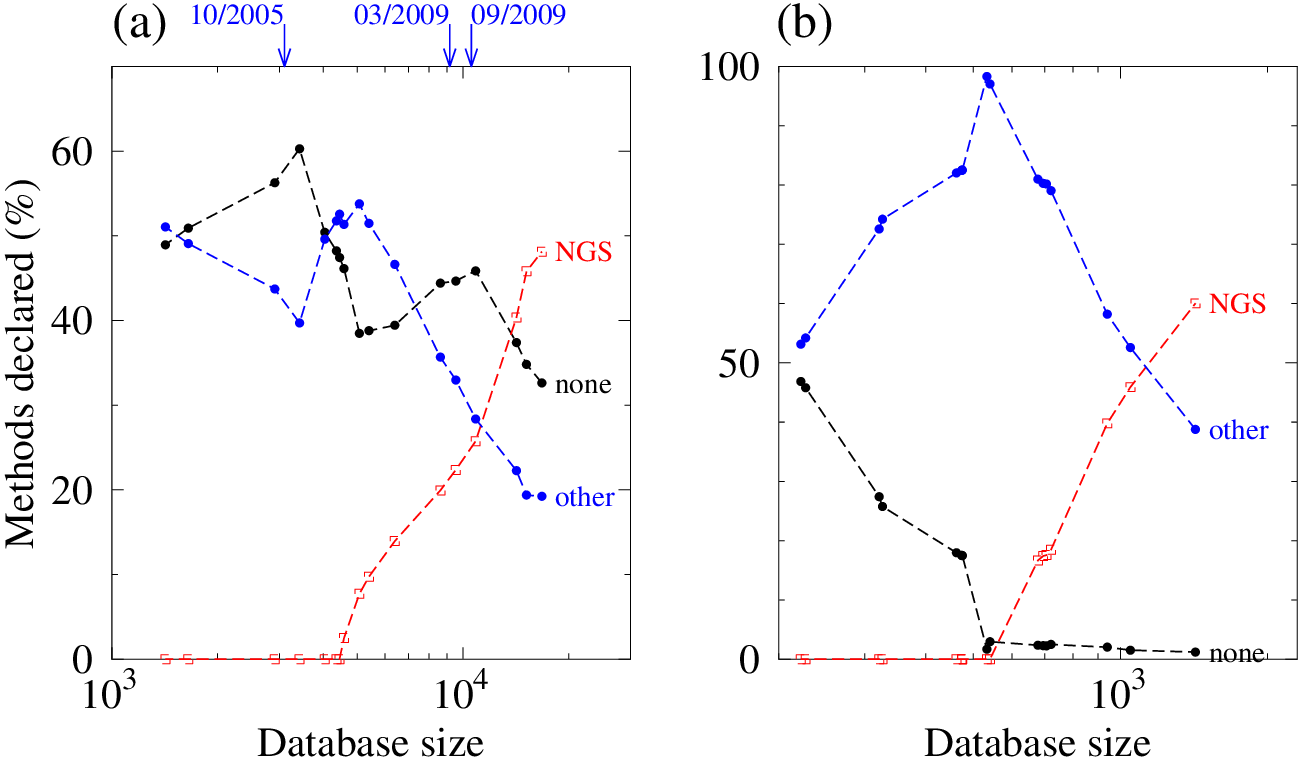}
\end{center}
\caption{
{\bf Number of experimental methods declared in miRBase}.
Fractions are computed by counting the tags \texttt{experimental} declared for each miRNA.
Tag values '464', 'SOLiD' and 'Solexa' are collectively shown as red boxes 
refer to next-generation sequencing (NGS) methods. 
Blue boxes are all methods different from NGS (other) and black bullets are sequences for which no
method was given (none).
Part~(a) shows the complete database, and part~(b) the human miRNAs only.
Note that a large number of miRNAs are sequenced by several techniques, or resequenced with
different techniques at a later time.
Also, a large number of miRNAs have their experimental methods declared some time after their first appearance
in the database.
In other words, the annotation of methods is only approximately correlated with the appearance of new miRNAs in the
database.
}
\label{fig-methods}
\end{figure*}

The results presented here let us to the conclusion that the sudden change in network topology between version~13 and~14
is not dominated by a change in similarity scores, although these scores do decrease continually since version 7.
They are also not present in the largest species group.
Clearly, there must have happened an important technological or methodological change in the way new miRNAs were
discovered around the year 2009.
Until 2005, the detection of miRNAs was usually preceded by an extensive bioinformatics analysis of known sequences.
For example, some techniques would use conserved stem loops of precursor miRNAs~\cite{lim03,lai03c}
or phylogenetic shadowing~\cite{berezikov05,berezikov06} for identifying new miRNAs.
These techniques rely heavily on sequence alignments of known miRNAs and this appears to be reflected 
by the steady increase of clustering coefficients up to 2005 seen in Figs.~\ref{fig-mature-12} and \ref{fig-pre-12}.
For human miRNA the situation is less clear as shown in Fig.~\ref{fig-mature-hs-12}, there is an initial steady decrease
until version 6, then a steep increase for version 7. 
From 2005 onwards clustering coefficients decrease for all situations which coincides with the onset
of RNA-seq, high throughput deep sequencing techniques widely used in studies of the transcriptome. 
This is shown in Figure~\ref{fig-methods} which presents the evolution of sequencing methods declared in the database
over time.
Differently from previous techniques, RNA-seq
has some features that are responsible for the abundance of data generated by this technique:
cDNA molecules are sequenced in parallel, producing large amounts of sequence data and the
identification of sequences (like miRNAs) can be made without prior sequencing knowledge~\cite{snyder09}. 
In 2009 there appears to be deluge of RNA-seq data that completely changes the network topology.
The computational methods to identify candidate miRNAs from the deep sequencing data operate under a very different
set of requirements than earlier methods.
In particular methods such as miRDeep~\cite{friedlander08} explicitly avoid cross-species comparisons.
Several more recent methods of miRNA discovery rely less on similarities with know miRNAs and employ other strategies
such as biochemical characteristics of miRNA biogenesis~\cite{hendrix10} or try to predict Drosha processing sites to improve miRNA
prediction~\cite{helvik07}.
This may explain the steady decrease in clustering coefficients starting in 2005 and perhaps the sharp
drop seen in 2009.
For human miRNA on the other hand, where much more effort was spent in the early days of miRNA sequencing, 
the database appears to be much more complete and the drastic drop around version 13 and 14 is absent.

\section*{Conclusions}

We developed a procedure which makes use of sequence similarities
to evaluate if network topologies could be biased by network growth.
We applied the technique to the network topology analysis of the chronological history of sequences deposited in
miRBase.
We were able to show that the network topology, notably the clustering coefficients, 
shows a clear database construction bias.
This means the resulting network topology depends critically at which point of time in the history of miRBase
it was performed.
For example, an analysis performed in 2009 would arrive at totally different network properties 
than the same analysis made a year earlier.
We believe that this substantiates some of the criticism that indiscriminate interpretation of network topology 
has received recently~\cite{hakes08,lima09}.


\section*{Author contribution}
    
GBS wrote the Perl scripts of the sampling simulations, 
CPSG performed the initial analysis on the network topology,
DFL provided the biological analysis of the miRNA methods, 
ACS and GW provided conceptual advice and supervised GBS, CPSG and DFL,
and wrote the paper.

\section*{Acknowledgements}

Funding: CNPq, Capes, Fapemig and National Institute for Science and Technology of
Complex Systems.

\bibliography{complete-gbc,redes}
\bibliographystyle{nature}

\end{document}